\documentstyle[12pt] {article}
\newcommand{\beq}{\begin{equation}}
\newcommand{\eeq}{\end{equation}}
\newcommand{\beqa}{\begin{eqnarray}}
\newcommand{\eeqa}{\end{eqnarray}}
\newcommand{\ba}{\begin{array}}
\newcommand{\ea}{\end{array}}

\begin{document}
{\large \bf On the Limit Cycle of an Inflationary Universe} 

\vskip 1. truecm

\begin{center}
{\bf Luca Salasnich}\footnote{Electronic address: 
salasnich@math.unipd.it}
\vskip 0.5 truecm
Dipartimento di Matematica Pura ed Applicata \\
Universit\`a di Padova, Via Marzolo 8, I 35131 Padova, Italy \\
and\\
Istituto Nazionale di Fisica Nucleare, Sezione di Padova,\\
Via Marzolo 8, I 35131 Padova, Italy
\end{center}

\vskip 1. truecm

\begin{center}
{\bf Abstract}
\end{center}

\vskip 0.5 truecm
\par
We study the dynamics of a scalar inflaton field 
with a symmetric double--well potential and prove rigorously 
the existence of a limit cycle in its phase space. 
By using analytical and numerical arguments we show that 
the limit cycle is stable and give an analytical formula for its period. 

 
\newpage

\par
The nonlinear and chaotic behaviour of classical field theories is currently 
subject of intensive research [1--3] and, in this respect, it is of great 
interest to investigate the existence and properties of limit cycles, 
which are inherently nonlinear phenomena [4,5]. 
\par
In a previous paper [6] we studied the stability of a scalar 
inflaton field with a symmetric double--well self energy. 
We showed that the value of the inflaton field in the vacuum 
is a bifurcation parameter 
which changes the phase space structure and that  
for some functional solutions of the Hubble "constant" 
the system goes to a limit cycle, i.e. to a periodic orbit. 
\par
In this paper we analyze the properties of this limit cycle 
by using analytical and numerical arguments. We show that 
the limit cycle is unique and stable and 
give an analytical formula for its period. 
\par
To solve the three major cosmological problems, i.e. the flatness problem, the 
homogeneity problem, and the formation of structure problem, 
it is generally postulated that the universe, at a very early stage after 
the big bang, exhibited a short period of exponential expansion, the 
so--called inflationary phase [7--10]. 
All the inflationary models assume the existence of a scalar field 
$\phi$, the so--called inflation field, with Lagrangian [9,11] 
\beq
L={1\over 2}\partial_{\mu}\phi \partial^{\mu}\phi -V(\phi )
\eeq
where the potential $V(\phi )$ depends on the type of inflation model 
considered. Here we choose a real field but also 
complex scalar can be used [9]. The scalar field, 
if minimally coupled to gravity, satisfies the equation 
\beq
\Box \phi = {\ddot \phi} + 3 \big({\dot a\over a}\big) 
{\dot \phi} - {1\over a^2} \nabla^2 \phi 
= -{\partial V \over \partial \phi},
\eeq
where $\Box$ is the covariant d'Alembertian operator and $a$ is 
the cosmological scale factor. 
The parameter $G$ is the gravitational constant 
($G=M_p^{-2}$ with $\hbar =c=1$ and $M_p =1.2 \cdot 10^{19}$ GeV 
the Plank mass) and $H={\dot a}/a$ is the Hubble "constant", 
which in general is a function of time (Hubble function). 
We suppose that in the universe there is only the inflaton field,  
so the Hubble function $H$ is related to the energy density of 
the field by
\beq
H^2 +{k\over a^2}=
\big({{\dot a} \over a}\big)^2+{k\over a^2}=
{8 \pi G \over 3} \big[
{{\dot \phi}^2 \over 2}+{(\nabla \phi )^2\over 2}+ V(\phi ) 
\big].
\eeq 
Immediately after the onset of inflation, the cosmological scale factor 
$a$ grows exponentially [9]. Thus the term $\nabla^2 \phi /a^2$ is 
generally believed to be negligible and, if the inflaton field is 
sufficiently uniform (i.e. ${\dot \phi}^2$, $(\nabla \phi )^2 << V(\phi )$), 
we end up with a classical nonlinear scalar field theory in one dimension
\beq
{\ddot \phi } + 3 H(\phi ) {\dot \phi} + 
{\partial V \over \partial \phi}=0,  
\eeq
where the Hubble function $H$ satisfies the equation
\beq
H^2 = {8 \pi G \over 3} V(\phi ).
\eeq
\par
The potential $V(\phi )$ depends on the type of inflation model considered 
and we choose a symmetric double--well potential
\beq
V(\phi )={\lambda \over 4}(\phi^2 -v^2)^2 ,
\eeq
where $\pm v$ are the values of the inflaton field in the vacuum, 
i.e. the points of minimal energy of the system. 
\par
One of the main difficulties in constructing models with potentials 
suitable for inflation is that these potentials must be flat enough to allow 
a sufficiently long period of inflation [8,9]. In this respect our model 
is very schematic but it can be seen as a toy model for classical 
nonlinear dynamics with the attractive feature that it emerges 
form the inflationary cosmology. Obviously the complete study of the 
dynamics of the inflaton may be addressed only by a complete quantum 
field theory approach able to predict not only 
the behaviour of the classical value of the inflaton field 
but also the associated quantum fluctuations. However 
the quantization of the inflation scenario is still an open problem [10] 
(an interesting stochastic approach can be found in [12]). 
\par
In the paper [6] we showed that the inflaton field value 
in the vacuum $v$ is a bifurcation parameter. If $v=0$ in the phase space 
there is only one stable fixed point $(\phi =0,{\dot \phi}=0)$, which is 
an attractor. Instead for $v\neq 0$ there are three fixed points: 
$(\phi =0,{\dot \phi} =0)$, which is unstable, 
and $(\phi = \pm v,{\dot \phi} =0)$, which are stable. 
\par
The Hubble function is determined by solving the equation (5). There 
are four possible continuous solutions : 
\beq
H(\phi )= \pm \gamma |\phi^2-v^2| , 
\eeq
but also: 
\beq
H(\phi )= \pm \gamma (\phi^2-v^2) , 
\eeq
where $\gamma=\sqrt{2\pi G\lambda / 3}$ is the friction parameter. 
The choice of the solution is crucial for the dynamical 
evolution of the system. 
\par
By using the Bendixon criterion [13] (discussed in detail in [6]) we obtain 
that if $H(\phi )=\gamma |\phi^2-v^2|$ then the Hubble function does not 
change sign and we do not find periodic orbits. The 4th--order Runge--Kutta 
numerical integration [14] of the equations of motion shows that for 
$v\neq 0$ the inflation field approaches one of its two stable fixed point 
attractors, and that the Hubble function goes to zero with an oscillatory 
behaviour (see Figure 1). 
Instead, if we choose $H(\phi )=\gamma (\phi^2-v^2)$ then the Hubble 
function can change sign and the Bendixon criterion admits for $v\neq 0$ 
the existence of a limit cycle. In fact, the numerical 
calculations plotted in Figure 2 show that a limit cycle exists and that 
the Hubble function oscillates forever. 
\par
Now we want analyze in detail the properties of this limit cycle. 
The equation of motion of the inflaton field with 
$H(\phi )=\gamma (\phi^2-v^2)$ reads 
\beq
{\dot \phi}+ 3 \gamma (\phi^2 - v^2) {\dot \phi}+ 
\lambda \phi ( \phi^2 - v^2 ) = 0 \; .
\eeq
This equation can be written as
\beq
{d\over dt}\big[ {\dot \phi} + 3 \gamma \int_0^{\phi}(u^2-v^2) du \big] 
+ \lambda \phi (\phi^2 - v^2) = 0 \; ,
\eeq
and if we put
\beq
F(\phi )= 3 \int_0^{\phi}(u^2-v^2) du = \phi (\phi^2 - 3 v^2) \; , 
\;\;\;\; G(\phi ) = \phi (\phi^2 - v^2 ) \; ,
\eeq
and also $\omega = {\dot \phi} + \gamma F(\phi )$, we obtain the system
\beq
\left\{ 
\ba{ccc}
{\dot \phi} & = & \omega - \gamma F(\phi ) 
\\
{\dot \omega} & = & - \lambda G(\phi ) \; . 
\ea 
\right. 
\eeq 
For systems of this kind the Lienard theorem [15,16] states that there 
is an unique and stable limit cycle if the following conditions are satisfied: 
$F(\phi )$ is an odd function and $F(\phi )=0$ only at $\phi =0$ and 
$\phi = \pm \alpha $; $F(\phi ) < 0$ for $0< \phi < \alpha $, 
$F(\phi ) > 0$ and is increasing for $\phi > \alpha $; $G(\phi )$ is an 
odd function and $\phi G(\phi ) >0 $ for all 
$\phi > \alpha$. It is easy to check that the functions $F(\phi )$ and 
$G(\phi )$ defined by (11) satisfy all the conditions of the Lienard 
theorem with $\alpha = v$. The cubic force $G(\phi )$ tends to reduce any 
displacement for large $|\phi |$, whereas the damping $F(\phi )$ is 
negative at small $|\phi |$ and positive at large $|\phi |$. 
Since small oscillations are pumped up and large oscillations are damped down, 
it is not surprising that the system tends to seattle into a self--sustained 
oscillation of some intermediate amplitude. 
\par
Figure 2 and Figure 3 show that both internal and external initial conditions 
generate trajectories which approach the limit cycle, so we have also 
a numerical evidence of the stability of the limit cycle. 
\par
Let us consider a typical trajectory of the Lienard system (12). 
After the scaling $\psi = \lambda \omega $ we obtain 
\beq
\left \{ 
\ba{ccc}
{\dot \phi} & = & \lambda [\psi - {\gamma\over \lambda }F(\phi )]  \; , 
\\
{\dot \psi} & = & - G(\phi ) \; .
\ea 
\right. 
\eeq
The cubic nullcline $\psi = (\gamma /\lambda ) F(\phi ) $ is the key to 
understand the motion [5]. Suppose that $\lambda >>1$ and 
the initial condition is far from 
the cubic nullcline, then (13) implies 
$|{\dot \phi }|\sim O(\lambda ) >> 1$; hence the velocity 
is enormous in the horizontal direction and tiny in the vertical direction, 
so trajectories move practically horizontally. If the initial condition 
is above the nullcline then ${\dot \phi} >0$, thus the trajectory moves 
sideways toward the nullcline. However, once the trajectory 
gets so close that 
$\psi \simeq (\lambda /\gamma) F(\phi )$ then 
the trajectory crosses the nullcline vertically and moves slowing along 
the backside of the branch until it reaches the knee and can jump sideways 
again. The period $T$ of the limit cycle is essentially the time required 
to travel along the two slow branches, since the time spent in the jumps is 
negligible for large $\lambda$. 
By symmetry, the time spent on each branch is the same so we have
\beq
T \simeq 2 \int_{t_A}^{t_B} dt \; 
\eeq
where $A$ and $B$ are the initial and final points on the positive slow branch. 
To derive an expression for $dt$ we note that on the slow branches with 
a good approximation $\psi \simeq (\gamma /\lambda ) F(\phi )$ and thus 
\beq
{d\psi \over dt} \simeq {\gamma \over \lambda } F'(\phi ) {d\phi \over dt} 
= 3 {\gamma \over \lambda} (\phi^2 - v^2){d\phi \over dt} \; .
\eeq
Since from (13) $d\psi / dt = - \phi (\phi^2 -v^2)$, we obtain
\beq
dt \simeq - 3 {\gamma\over \lambda } {d\phi \over \phi } \; ,
\eeq
on the slow branches. The slow positive branch begins at 
$\phi_A = 2\gamma v/\lambda$ and ends at $\phi_B=\gamma v/\lambda$, hence 
\beq
T\simeq 2 \int_{t_A}^{t_B} dt \simeq - 6 {\gamma \over \lambda} 
\int_{\phi_A}^{\phi_B} {d\phi \over \phi} 
\simeq 6 {\gamma \over \lambda } \ln{2} \; .
\eeq
Because $\gamma = \sqrt{2\pi G \lambda /3}$ we have
\beq
T\simeq 2 \ln{2} \sqrt{6\pi G\over \lambda} \; .
\eeq
Note that the period is $v$ independent. 
\par
In summary, we have proved the existence 
and stability of a limit cycle in the phase space of a scalar 
inflaton field $\phi$ with a symmetric 
double--well potential $V(\phi )$ and a friction term in the 
equation of motion proportional to $V(\phi )$. Then we have obtained 
an analytical estimation of the period of the limit cycle. 

\vskip 0.5 truecm

\begin{center}
*****
\end{center}
\par
The author is greatly indebted to V.R. Manfredi and M. Robnik 
for many enlightening discussions. 

\newpage

\section*{References}

\begin{description}

\item{\ [1]} T. Kawabe and S. Ohta, Phys. Lett. B {\bf 334}, 127 (1994); 
T. Kawabe, Phys. Lett. B {\bf 343}, 225 (1995)

\item{\ [2]} L. Salasnich, Phys. Rev. D. {\bf 52}, 6189 (1995); 
S. Graffi, V. R. Manfredi, L. Salasnich, Mod. Phys. Lett. B {\bf 7}, 747 
(1995)

\item{\ [3]} J. Segar and M. S. Sriram, Phys. Rev. D {\bf 53}, 3976 (1996)

\item{\ [4]} A. H. Nayfeh and B. Balachandran, {\it Applied Nonlinear Dynamics} 
(J. Wiley, New York, 1995)

\item{\ [5]} M. Farkas, {\it Periodic Motion} (Springer, Berlin, 1994)

\item{\ [6]} L. Salasnich, Mod. Phys. Lett. A {\bf 10}, 3119 (1995)

\item{\ [7]} A. H. Guth, Phys. Rev. D {\bf 23} B, 347 (1981)

\item{\ [8]} A. D. Linde, Phys. Lett. B {\bf 108}, 389 (1982); 
A. D. Linde, Phys. Lett. B {\bf 129}, 177 (1983)

\item{\ [9]} A. D. Linde, {\it Particle Physics and Inflationary Cosmology} 
(Harwood Academic Publishers, London, 1988)

\item{\ [10]} R. H. Brandenberger, 
in SUSSP Proceedings {\it Physics of the Early 
Universe}, Eds. J.A. Peacock, A. F. Heavens and A. T. Daves (Institute of 
Physics Publishing, Bristol, 1990)

\item{\ [11]} C. Itzykson and J. B. Zuber, {\it Quantum Field Theory} 
(McGraw--Hill, New York, 1985)

\item{\ [12]} C. Beck, Nonlinearity {\bf 8}, 423 (1995)

\item{\ [13]} I. Bendixson, Acta Math. {\bf 24}, 1 (1901)

\item{\ [14]} Subroutine D02BAF, The NAG Fortran Library, Mark 14, Oxford: NAG 
Ltd. and USA: NAG Inc. (1990)

\item{\ [15]} A. Lienard, Rev. Gen. Electr. {\bf 23}, 901 (1928)

\item{\ [16]} D. W. Jordan and P. Smith, {\it Nonlinear Ordinary Differential 
Equations} (Oxford Univ. Press, Oxford, 1987)

\end{description}

\newpage
 
\parindent=0.pt
 
\section*{Figure Captions} 
\vspace{0.6 cm}
 
{\bf Figure 1}: The Hubble function {\it vs} time (up) and the phase space 
trajectory of the inflaton field (down); for $H(\phi )= 
\gamma |\phi^2 - v^2 |$ with $\gamma =1/2$, $\lambda =3$ and $v=1$. 
Initial conditions: $\phi =0$ and ${\dot \phi}=1/2$. 

{\bf Figure 2}: The Hubble function {\it vs} time (up) and the phase space 
trajectory of the inflaton field (down); for $H(\phi )= 
\gamma (\phi^2 - v^2 )$ with $\gamma =1/2$, $\lambda =3$ and $v=1$. 
Initial conditions: $\phi =0$ and ${\dot \phi}=4$. 

{\bf Figure 3}: The Hubble function {\it vs} time (up) and the phase space 
trajectory of the inflaton field (down); for $H(\phi )= 
\gamma (\phi^2 - v^2 )$ with $\gamma =1/2$, $\lambda =3$ and $v=1$. 
Initial conditions: $\phi = -1/2$ and ${\dot \phi}=0$. 

\end{document}